\input harvmac.tex

%
%

\nref
\CremmerI{ E. Cremmer, B Julia and J. Scherk, ``Supergravity Theory in 11
Dimensions," Physics Letters Volume 76B, Number 4 ( 409 - 412 ) }

\nref
\DasguptaI{K. Dasgupta and S. Mukhi, ``Orbifolds of M-Theory,"
hep-th/9512196; TIFR/TH/95-64 }

\nref
\DuffI{M.J. Duff, J.T. Jiu, R. Minasian, ``Eleven Dimensional Origin of
String/String Duality : A One Loop Test," hep-th/9506126, CTP-TAMU-26/95 }

\nref
\GinspargI{ P. Ginsparg, ``On Torodial Compactification of Heterotic
Superstrings," Physical Review D Volume 35, Number 2 ( 648 - 654 ) }

\nref
\GriffithsI{ P. Griffiths and J. Harris, {\it Principles of Algebraic Geometry}
John Wiley and Sons (See particularly around page 590) }

\nref
\HullI{C.M. Hull and P.K. Townsend,``Unity of Superstring Dualities,"
hep-th/9410167; QMW-94-30; R/94/33 }

\nref
\HuqI{M. Huq and M.A. Namazie, ``Kaluza-Klein Supergravity in Ten
Dimensions," Classical and Quantum Gravity {\bf 2} ( 1985 ) 293-308 }

\nref
\SchwarzI{J.H. Schwarz, ``M-Theory Extensions of T-Duality,"
hep-th/9601077, CALT-68-2034. }

\nref
\SeibergI{M. Dine, P. Huet and N. Seiberg, ``Large and Small Radius in
String Theory," Nuclear Physics B {\bf 322} ( 1989 ) 301-316 }

\nref
\SenI{A. Sen, ``String String Duality Conjecture in Six Dimensions and Charged
Solitonic Strings," hep-th/9504027, TIFR-TH-95-16 }

\nref
\WaldI{R.M. Wald, ( 1984 ) {\it General Relativity} Chicago : The University of
Chicago Press}

\nref
\WessI{J. Bagger and J. Wess, ( 1992) {\it Supersymmetry and Supergravity}
Princeton : Princeton University Press }

\nref
\WittenI{E. Witten, ``String Theory Dynamics in Various Dimensions,"
hep-th/9503124; IASSNS-HEP-95-18 }

\nref
\WittenII{N. Seiberg and E. Witten, ``Electric-Magnetic Duality, Monople
Condensation, and Confinement in N=2 Supersymmetric Yang-Mills Theory,"
hep-th/9407087; RU-94-52; IAS-94-43 }

\nref
\WittenIII{P. Horava and E. Witten, ``Heterotic and Type I String
Dynamics from Eleven Dimensions," hep-th/9510209; IASSNS-HEP-95-86;
PUPT-1571 }

\nref
\WittenVI{M.B. Green, J.H. Schwarz and E. Witten, ( 1987 ) {\it Superstring
Theory: Loops Amplitudes, Anomalies and Phenomenology }
Cambridge : Cambridge University Press }

\nref
\WittenVII{ E. Witten, ``Five-Branes and M-Theory on an Orbifold,"
hep-th/9512219; IASSNS-HEP-96-?? }

%
%

\Title{
	\vbox{
		\baselineskip12pt
		\hbox{RU 96-xx}
		\hbox{hep-th/9601102}
		\hbox{Rutgers Theory}
		}
	}
	{
	\vbox{
		\centerline{M-Theory}
		\vskip2pt
		\centerline{ and }
		\vskip2pt
		\centerline{String-String Duality}
		}
	}

\centerline{
  Kelly Jay Davis\foot{ With ``Natural Ingredients" from Luscious Jackson}
}
\bigskip
\centerline{Rutgers University}
\centerline{Serin Physics Laboratory}
\centerline{Piscataway, NJ 08855}

%
%

\vskip .3in
\centerline{ABSTRACT}
In this article we  examine the compatibility of some recent results,
results relating M-Theory to String Theory, with the string-string duality
conjecture in six-dimensions. In particular, we rederive the relation
between M-Theory and Type IIA strings. We then go on to examine in detail
M-Theory on $K3 \times S^{1}$ and its relation to the Heterotic theory on
$T^{4}$. We conclude with some remarks on M-Theory on $T^{4}\times(S^{1} /
{\bf Z}_{2})$ and its relation to the Type II theory on $K3$.
\Date{1/19/96}

%
%

\newsec{Introduction}

In the past year much has happened in the field of string theory.
Old results relating the two Type II string theories \SeibergI\ and the
two Heterotic string theories \GinspargI\ have been combined with newer
results relating the Type II theory and the Heterotic theory \WittenI
\SenI\  as well as the Type I theory and the Heterotic theory \WittenIII\
to obtain a single ``String Theory." In addition, there has been much
recent progress in interpreting some, if not all, properties of String
Theory in terms of an eleven-dimensional M-Theory \DasguptaI  \HullI
\SchwarzI \WittenI \WittenIII. In this paper we will perform a
self-consistency check on the various relations between M-Theory and
String Theory. In particular, we will examine the relation between String
Theory and M-Theory by examining its consistency with the string-string
duality conjecture of six-dimensional String Theory. So, let us now take a
quick look at the relations between M-Theory and String Theory some of
which we will be employing in this article.

In Witten's paper \WittenI\ he established that the strong
coupling limit of Type IIA string theory in ten-dimensions
is equivalent to eleven-dimensional supergravity \CremmerI
\WittenVI\ on a ``large" $S^1$. As the low energy limit of
M-theory is eleven-dimensional supergravity \WittenIII, this
relation states that the strong coupling limit of Type IIA
string theory in ten-dimensions is equivalent to the
low-energy limit of M-Theory on a ``large" $S^1$. In the
paper of Witten and Horava \WittenIII, they establish that
the strong coupling limit of the ten-dimensional
$E_{8} \times E_{8}$ Heterotic string theory is equivalent
to M-Theory on a ``large" $S^{1} / {\bf Z}_{2}$. Recently,
Witten \WittenVII , motivated by Dasgupta and Mukhi \DasguptaI,
examined M-Theory on a ${\bf Z}_{2}$ orbifold of the
five-tours and established a relation between M-Theory on
this orbifold and Type IIB string theory on $K3$. Also,
Schwarz \SchwarzI\ very recently looked at M-Theory and
its relation to T-Duality.

As stated above, M-Theory on a ``large"
$S^1$ is equivalent to a strongly coupled
Type IIA string theory in ten-dimensions. Also, M-theory on a ``large" $S^{1} /
{\bf Z}_{2}$ is equivalent to a strongly coupled $E_{8} \times E_{8}$ Heterotic
string theory in ten-dimensions. However, the string-string duality
conjecture in six dimensions states that the strongly coupled limit of a
Heterotic string theory in six-dimensions on a four-torus is equivalent
to a weakly coupled Type II string theory
in six-dimensions on $K3$. Similarly, it states that the strongly coupled
limit of a Type II theory in six dimensions on $K3$ is equivalent
to a weakly coupled Heterotic string theory in six-dimensions on a
four-torus. Now, as a strongly coupled Type IIA string theory in ten-dimensions
is equivalent to the low energy limit
of M-Theory on a ``large" $S^1$, the low energy
limit of M-Theory on $S^{1} \times K3$ should be equivalent to a
weakly coupled Heterotic string theory on a four-torus by way
of six-dimensional string-string duality. Similarly,
as a strongly coupled $E_{8} \times
E_{8}$ Heterotic string theory in  ten-dimensions
is equivalent to the low energy
limit of M-Theory on a ``large" $S^{1} / {\bf Z}_{2}$,
the low energy limit of M-Theory
on $S^{1} / {\bf Z}_{2} \times T^{4}$ should be equivalent to a weakly
coupled Type II string theory on $K3$.  The first of the above two consistency
checks on the relation between M-Theory and String
Theory will be the subject of
this article. However, we will comment on the second consistency check in
our conclusion.

%
%

\newsec{M-Theory $\sim$ Type IIA Equivalence}

In this section we will recount Witten's results \WittenI \WittenIII\
establishing
the equivalence between a strongly coupled Type IIA string theory in
ten-dimensions and the low-energy limit of M-Theory on a ``large" $S^{1}$. To
do so we will rely heavily upon the stability of
BPS saturated states. Hence, we will
first briefly review the reasoning behind such
stability in a four-dimensional case
before we go to ten-dimensions.

%
%

\subsec{ Stability of BPS Saturated States }

In this subsection we will derive the stability of BPS saturated states in a
four-dimensional example which gives the
flavor and motivation behind the stability
relied upon later in this section. Consider a $N = 2$ supersymmetric theory in
four-dimensions. Such a theory possess supercharges $Q^{L}_{\alpha}$ where
$\alpha$ is a Weyl spinor index and $L = 1 , 2$. Such supercharges satisfy the
following algebraic relations,
\eqn
\AntiCommutationRelationsI{
\eqalign{
& \{ Q_{\alpha}^{L} , {\bar Q}_{ {\dot \alpha} M} \}  =
2 \sigma_{\alpha {\dot \alpha} }^{m}P_{m} \delta^{L}_{M} \cr
& \{ Q_{\alpha}^{L} , Q_{\beta}^{M} \}  =
\epsilon_{\alpha \beta} Z^{ \langle L, M \rangle } , \cr
}
}
\noindent where we are employing the standard notation of Wess and Bagger
\WessI.  Now, following \WessI, one can consider a particle of mass $M$ in its
rest frame. Such a particle has momentum $P= ( -M, 0 )$ which implies that the
above relations take the form,
\eqn
\AntiCommutationRelationsII{
\eqalign{
& \{ Q_{\alpha}^{L} , {(Q_{\beta}^{M})}^{\dagger} \}  =
2 M \delta^{L}_{M} \delta^{\beta}_{\alpha} \cr
& \{ Q_{\alpha}^{L} , Q_{\beta}^{M} \}  =
\epsilon_{\alpha \beta} Z^{ \langle L, M \rangle } .\cr
}
}
\noindent  As $Z^{LM}$ is anti-symmetric and
$L \& M = 1 , 2$, one has $Z^{LM} = Z
\epsilon^{LM}$. So, this implies,
\eqn
\AntiCommutationRelationsIII{
\eqalign{
& \{ Q_{\alpha}^{L} , {(Q_{\beta}^{M})}^{\dagger} \}  =
2 M \delta^{L}_{M} \delta^{\beta}_{\alpha} \cr
& \{ Q_{\alpha}^{L} , Q_{\beta}^{M} \}  =
\epsilon_{\alpha \beta} \epsilon^{L M} Z .\cr
}
}
\noindent Next, we may define two new supercharges $q^{L}_{\alpha}$
which are linear combinations of the $Q^{L}_{\alpha}$,
\eqn
\NewSuperChargesI{
\eqalign{
& q^{1}_{\alpha}  = { { 1 } \over { \sqrt{2} } } ( Q^{1}_{\alpha} +
\epsilon_{\alpha \beta} {(Q_{\beta}^{2})}^{\dagger}  ) \cr
& q^{2}_{\alpha}  = {{1} \over { \sqrt{2} } } ( Q^{1}_{\alpha} -
\epsilon_{\alpha \beta} {(Q_{\beta}^{2})}^{\dagger}  ) .\cr
}
}
\noindent Finally, by way of \AntiCommutationRelationsIII,
these operators satisfy the following anti-commutation
relations,
\eqn
\AntiCommutationRelationsIV{
\eqalign{
& \{ q_{\alpha}^{1} , {(q_{\beta}^{1})}^{\dagger} \}  =
\delta^{\beta}_{\alpha} ( 2 M  + Z )\cr
& \{ q_{\alpha}^{2} , {(q_{\beta}^{2})}^{\dagger} \}  =
\delta^{\beta}_{\alpha} ( 2M  - Z) . \cr
}
}
Now, the point to notice about all of this
algebra is that the second equation in
\AntiCommutationRelationsIV\ has positive semi-definite
eigenvalues. This implies $M \ge {Z /  2}$. Hence, there
is a lower bound on the mass of a state in a
representation of the supersymmetry algebra with central
charge $Z$, and this lower bound  is $Z / 2$. States which saturate
this bound, particles with mass $Z / 2$,
are said to be BPS saturated states and it is these states which
are ``stable" as we will now proceed to show.

A question which may now arise : Physically, to what does the central charge
correspond? The answer depends upon the particular $N = 2$ theory one is
considering. So, we will just touch upon a
single example from the work of Witten
and Seiberg \WittenII. In this paper they considered a $N = 2$ Yang-Mills
theory in four-dimensions with gauge group $SU(2)$. In their example
the gauge group $SU(2)$ was spontaneously broken to $U(1)$.
So, any particular particle of this spontaneously broken gauge theory possess a
$U(1)$ electric charge $n_{e}$ and a $U(1)$ magnetic charge $n_{m}$,
where we can always choose units such that
$n_{e}$ and $n_{m}$ are integers. The
electric and magnetic charges are related to the central charge $Z$
by way of two complex numbers $a$ and $a_{D}$ which arise in the theory. The
complex numbers $a$ and $a_{D}$ allow us to define a third complex number
${\bf Z} = n_{e} a + n_{m} a_{D}$ associated with the electric and magnetic
charges of a particular particle. For this
particular particle, with electric charge
$n_{e}$ and magnetic charge $n_{m}$, the central
charge for its representation of the
supersymmetry algebra is given by $Z = 2 \sqrt{2} | {\bf Z}  |$,
where $| {\bf Z}  |$ is the modulus of the complex number ${\bf Z}$.
So, this particular particle has mass $M \ge  \sqrt{2} | {\bf Z}  |$.

Now let us consider a particular particle which is BPS saturated. In other
words, let us consider a particle with
charges $n_{e}$ and $n_{m}$ and mass $M$
such that $M = \sqrt{2} | {\bf Z}  |$. Is such a
particle stable? Let us consider the
possible modes of decay for such a particle. In considering such a decay let us
assume  that the ratio $a / a_{D} $ is not real. Now, let us assume that
the state $Z$ may decay into several other states $Z_{i}$ with electric charges
$n_{e,i}$ and magnetic charges $n_{m,i}$ as well as charge vectors
${\bf Z}_{i} = n_{e,i} a + n_{m,i} a_{D}$. As the ratio
$a / a_{D}$ is, by assumption, not real, charge conservation implies that
${\bf Z} = \sum {\bf Z}_{i}$. Furthermore, the triangle inequality
implies that $| {\bf Z} | \le \sum | {\bf Z}_{i} |$.
Employing the inequality $M_{i} \ge \sqrt{2}  | {\bf Z}_{i} |$  for each
state the triangle inequality implies $M \le \sum M_{i}$.
However, if the state $M$ is to decay into
the set of states $M_{i}$, then it is impossible that $M < \sum M_{i}$. So, this
implies that $M = \sum M_{i}$. This only occurs when all the ${\bf Z}_{i}$
are parallel to ${\bf Z}$. This is in turn only possible if $n_{e}$ and $n_{m}$
are not relatively prime; in other words, this
only occurs if there exist integers $q$,
$n$, and $m$ such that $n_{e} = q n$ and $n_{m} = q m$. In this case, the state
${\bf Z}$, among other things, can decay into $q$
BPS saturated particles each with
electric charge $n$ and magnetic charge $m$. Such states with
$n_{e}$ and $n_{m}$
not relatively prime are said to be neutrally stable.
However, one can see that if $n_{e}$ and $n_{m}$ are
relatively prime, then the state ${\bf Z}$  is stable against
decay as there exists no decay channel which preserves conservation
of momentum and conservation of charge.

Next let us consider a subset of our
previous example, as its analog  will appear in
the treatment of Type IIA string theory in
ten-dimensions. Again in this example
we will assume  the ratio $a / a_{D}$ is not real. However, in this case we will
consider a BPS saturated state with electric
charge $n_{e}$ and zero magnetic charge
$n_{m} = 0$. Assume it decays into a set of states $Z_{i}$ with electric
charges $n_{e,i}$ and magnetic charges $n_{m,i}$ as well as charge vectors
${\bf Z}_{i} = n_{e,i} a + n_{m,i} a_{D}$.
Charge conservation implies ${\bf Z} =
\sum {\bf Z}_{i}$, and the triangle inequality
implies $| {\bf Z} | \le \sum | {\bf Z}_{i} |$.
The BPS inequality  along with conservation of momentum implies $M = \sum
M_{i}$, and this only occurs when all the
${\bf Z}_{i}$ are parallel to ${\bf Z}$.
Hence, all the states $Z_{i}$ also only possess electric charge. Thus, as one
may easily see, a BPS saturated state
with only electric charge is only neutrally
stable, as a state with electric charge $n_{e}$
may decay into $n_{e}$ states, each
with electric charge $1$.

Now let us consider how the stability of states with relatively prime electric
$n_{e}$ and magnetic $n_m$ charges changes as the coupling constants of the
theory are varied. In particular, as we vary
the coupling constant $\Lambda$ of the theory
the only possible change in the above picture occurs when the dependence of
$a$ and $a_{D}$ upon $\Lambda$ is taken into account. So, we could consider
the hypothetical case\foot{ Note, this is not what occurs
in this theory, but we are using this as a hypothetical as
its analog does occur in Type IIA string theory
in ten-dimensions.} in which $a$ and $a_{D}$ both are
proportional to $\Lambda^{-1}$. In this case the
magnitude of $a$ and $a_{D}$ both vary as one varies $\Lambda$; however, the
ratio $a / a_{D}$ is constant for all $\Lambda$. So, in particular,
if the ratio is not
real for any given $\Lambda$, then it is not
real for all $\Lambda$ and the above
argument goes through for any coupling $\Lambda$. This implies that the
relatively prime BPS saturated states would exist
and be stable for all $\Lambda$!
But, if the dependence of $a$ and
$a_{D}$ upon $\Lambda$ is more complicated,
this may not be the case.

Again, as its analog will occur in the treatment of Type IIA string theory in
ten-dimensions, let us consider the neutral stability of BPS saturated states
with only an electric charge as $\Lambda$ is varied. As we showed above for
a given coupling, BPS saturated states with only an electric charge are
neutrally stable and may decay into a set of
BPS saturated states all with only electric charge.
Now, upon varying the coupling
constant $\Lambda$ of the theory, the continued
neutral stability of the electrically
charged BPS saturated states is dependent upon the $\Lambda$ dependence of
$a$ and $a_{D}$. If, hypothetically\foot{ Again, this
is not what occurs in this theory,
but we are using this as a hypothetical as it does occur
in Type IIA string theory in
ten dimensions.}, $a$ and $a_{D}$ were both
proportional to $\Lambda^{-1}$, then,
as above, the neutral stability of BPS saturated states
with only an electric charge
is guaranteed for all $\Lambda$. Again, if the
dependence of $a$ and $a_{D}$ upon
$\Lambda$ is more complicated, this may not be the case.
However, the simple dependence upon $\Lambda$ examined above will
indeed show up in the case we are really concerned with, Type IIA string
theory in ten-dimensions. So, we will be able
to use the neutral stability of the BPS
saturated states with only electric charge  in the next subsection to derive the
relation between the low-energy limit of M-Theory and the strong coupling
limit of Type IIA string theory in ten-dimensions.

%
%

\subsec{M-Theory $\sim$ Type IIA Equivalence}

In this subsection we will employ the
results of the previous subsection in deriving
the equivalence between a strongly coupled Type IIA string theory in
ten-dimensions and the low energy limit of M-Theory on a ``large" $S^{1}$. The
first step we will take in doing so is to examine Type IIA string theory in
ten-dimensions.

In ten-dimensions Type IIA string theory has
a low energy limit which is given by
Type IIA supergravity \WittenVI. So, let us
now look at Type IIA supergravity. Its
bosonic sector consists of a dilaton $\phi$,
a one-form $A$, a two-form $B$, a
three-form $A_{3}$, and a metric $g$.
The forms $A$, $B$, and $A_{3}$ give rise to
the field strengths $F = dA$, $H = dB$, and
$F_{4} = dA_{3}$, where $d$ denotes the
deRham operator in ten-dimensions. Also, in writing the bosonic portion of the
action, we will have need of the four-form
$F_{4}' = dA_{3} + A \wedge H$, where
$\wedge$ denotes the standard wedge product of forms.
The bosonic portion of the
Type IIA supergravity action in ten-dimensions
may be written as follows \WittenI\
( where we have taken $\alpha ' = 1$ ),
\eqn
\TypeIIASupergravityI{
	S = S_{NS} + S_{R}
}
\eqn
\TypeIIASupergravityII{
	S_{NS} =
	{1\over 2}
	\int d^{10}x\sqrt g e^{-2\phi}\left(R+4(\nabla\phi)^2 -
	{1\over 12}H^2\right)
}
\eqn
\TypeIIASupergravityIII{
	S_{R}=
	-\int d^{10}x \sqrt g\left({1\over 2\cdot 2!}F^2 +
	{1\over 2\cdot 4!}F_4'{}^2\right) -
	{1\over 4}\int F_4\wedge F_4\wedge B.
}
\noindent As the II of Type IIA denotes,
the full Type IIA theory possess a $N = 2$
supersymmetry. So, in particular, a generalization of
our previous remarks in the case
of four-dimensions holds in this case. The theory
has two supercharges $Q_{\alpha}$
and $Q_{\dot \alpha}'$, where $\alpha$ and $\dot \alpha$
denote Majorana-Weyl
spinor indices. These two supercharges, as in the four-dimensional case
\AntiCommutationRelationsI, satisfy anti-commutation relations of the general
form,
\eqn
\TypeIIAAntiCommutationRelationI{
	\{ Q , Q  \} \sim P
}
\eqn
\TypeIIAAntiCommutationRelationII{
	\{ Q' , Q' \} \sim P
}
\eqn
\TypeIIAAntiCommutationRelationIII{
	\{ Q_{\alpha} , Q_{\dot \alpha}' \} \sim \delta_{\alpha \dot \alpha} W,
}
\noindent where $W$ is a central charge.
Again, as was the case in four-dimensions,
this central charge leads to an inequality between the mass
$M$ of a particle and the
central charge $W$,
\eqn
\TypeIIAInequalityI{
	M \ge c_{0} W,
}
\noindent where $c_{0}$ is a ``constant" which,
as we shall show, only depends upon
the ten-dimensional string coupling constant $\lambda$. So, na\"\i vely it
looks as if the situation is exactly analogous
to that in four dimensions. However,
again, we should ask ourselves : Physically,
to what does the central charge
correspond?

In the four-dimensional case we considered a spontaneously broken Yang-Mills
theory with gauge group $SU(2)$ broken down to
$U(1)$ and we found the $U(1)$
electric and magnetic charges of a given
particle were related to the central charge of
that particle's supersymmetry algebra in a straight-forward manner.
However, in this situation we are dealing with supergravity;
hence, there is no obvious $U(1)$ gauge
group to which $W$ may be related. However, upon a
second look there is.

Looking  at \TypeIIASupergravityI\ one sees
that the bosonic field $A$ is invariant
under the $U(1)$ gauge transformation $\delta A = d \lambda_{0}$, where
$\lambda_{0}$ is a scalar. This symmetry also persists in the full Type IIA
supergravity theory. Hence, one might guess that charges with respect to this
$U(1)$ are related to the central charge $W$. We will find
that this is indeed the case. But to do so we must first
look at the relation between Type IIA supergravity
and eleven-dimensional supergravity.

A standard method of deriving Type IIA
supergravity is by dimensionally reducing
eleven-dimensional supergravity \HuqI \WittenVI. Dimensionally reducing
eleven-dimensional supergravity can be done by first putting the
eleven-dimensional
theory on $M \times S^1$, where $M$ is an arbitrary ten-dimensional manifold,
dropping the dependence of the eleven-dimensional supergravity fields on the
eleventh coordinate which parameterizes $S^1$, and decomposing the
eleven-dimensional supergravity fields in terms of representations of
$SO(1,9)$. In particular, the ten-dimensional one-form $A$ and the
ten-dimensional dilaton $\phi$ arise from the eleven-dimensional vielbein
as follows \WittenVI,
\eqn
\ElevenToTenSupergravityVielbein{
	e^{A}_{M}
	\rightarrow  \pmatrix{ e^{a}_{m} & A_{m}          \cr
					0   &    e^{ {2 \over 3} \phi }  \cr},
}
\noindent where the index $A$ is an
eleven-dimensional Lorentz index, $M$ is an
eleven-dimensional curved index, $a$ is a ten-dimensional Lorentz index, and
$m$ is a ten-dimensional curved index. The $10 \times 1$ block of zeros in
\ElevenToTenSupergravityVielbein\ can be obtained by making  a gauge
choice which uses up the Lorentz transformations between the first
ten-dimensions and the eleventh dimension.

Now, one may wonder how the gauge transformation $A = d \lambda_{0}$
is implemented from this eleven-dimensional view of Type IIA supergravity.
Consider a diffeomorphisim of the eleven-dimensional manifold $M \times S^1$
generated by an eleven-dimensional vector $V^{M}$ that does not depend upon
the eleventh coordinate of $M \times S^1$ and whose only non-zero component
is $V^{11}$. This diffeomorphisim is
obviously a symmetry of the eleven-dimensional
supergravity theory. However, one may wonder how it will be interpreted
in terms of $A_{m}$. To interpret the action of $V^{M}$ in terms of
$A_{m}$ let us first consider its action on $e^{A}_{M}$. One has the
standard relation,
\eqn
\IIDDiffeomorphisimI{
	e^{A}_{M} \rightarrow e^{A}_{M}  +
	V^{N} \partial_{N} e^{A}_{M} +
	e^{A}_{N} \partial_{M} V^{N},
}
\noindent where $\partial_{M}$ is the standard eleven-dimensional
derivative operator. Now, by definition $A_{m} = e^{11}_{m}$.
Hence, under this eleven-dimensional diffeomorphisim,
\eqn
\IIDDiffeomorphisimII{
	A_{m} \rightarrow A_{m} +
	e^{ {2 \over 3} \phi } \partial_{m} V^{11},
}
\noindent where we have employed the fact that the only non-zero
component of $V^{M}$ is $V^{11}$ and $A_{m}$ is independent
of the eleventh coordinate. One may now define a function
$\lambda_{0}$ as the solution to the differential equation
$\partial_{m} \lambda_{0} =
e^{ {2 \over 3} \phi } \partial_{m} V^{11}$. As $\phi$ and $V^{11}$ are both
ten-dimensional scalars, the solution $\lambda_{0}$ is also a
ten-dimensional scalar, and, noting equation
\IIDDiffeomorphisimII, $\lambda_{0}$
allows us to write the variation of $A_{m}$ under this eleven-dimensional
diffeomorphisim as $\delta A = d \lambda_{0}$. Hence, we obtain
the original gauge invariance of the Type IIA theory $\delta A =
d \lambda_{0}$ from an eleven-dimensional diffeomorphisim. But,
now we see that this gauge invariance simply comes from rotations of
$S^1$.

One may now ask oneself : What is electrically
charged with respect to this gauge
field $A$ from an eleven-dimensional point-of-view.
The answer to this question
is rather easy. Consider a plane-wave with a non-zero momentum $P_{11}$ in
the eleventh-dimension. Such a plane-wave is proportional to
$exp( i P_{11} x^{11} )$. Now a rigid rotation of the $S^{1}$ may be implemented
by the translation $x^{11} \rightarrow x^{11} + V^{11}$, where
$V^{11}$ is independent of $x^{11}$. Under such a rigid rotation the plane-wave
transforms as $exp( i P_{11} x^{11} ) \rightarrow exp( i P_{11} V^{11} )
exp( i P_{11} x^{11} )$. Hence, the plane-wave with momentum $P_{11}$ in the
eleventh-dimension transforms as if it has electric charge $P_{11}$ with
respect to the gauge field $A$. This is also true in general. A particle with
momentum $P_{11}$ in the eleventh-dimension will transform as if it has
electric charge  $P_{11}$ with respect to the gauge field $A$.

As we showed above, electrically charged objects from an eleven-dimensional
point-of-view are particles with a
non-zero momentum in the eleventh-dimension.
This has no obvious analog from a ten-dimensional point-of-view. However,
the U-Duality conjecture of Hull and Townsend \HullI\ implies
the existence of such electrically charged
objects in ten-dimensions. The U-Duality conjecture states that a
Type IIA string theory on a $d$-torus possess a discrete symmetry group,
with depends on the $d$ under consideration. This discrete symmetry
group, among other things, establishes a symmetry of Type IIA string theory
on a $d$-torus under the exchange of various electric and magnetic charges.
In particular, it postulates that there exists a symmetry of the toridially
compactified Type IIA string theory which
exchanges ``fundamental" electrically
charged objects, which we know to exist perturbatively, for instance winding
states, and objects which are electrically charged with respect
to the gauge field $A$ we have been
examining. Hence, if the U-Duality conjecture is true, then there exist
electrically charged particles with respect to the gauge
field $A$ in a toridially compactified Type IIA string theory. However,
one could consider taking all the radii of the torus to infinity to obtain a
ten-dimensional Type IIA string theory. Upon doing
so it would seem pathological if such electric charges with
respect to $A$ ceased to exist, as the gauge field $A$ continues to
exits in ten-dimensions in the same ``form" as it exists in $10 - d$
dimensions. Hence, motivated by  the
U-Duality conjecture, we will assume such electric charges
exist in ten-dimensional Type IIA string theory. But, we have not answered
the question: What objects are electrically charged with respect to this
gauge field $A$ from a ten-dimensional point-of-view?

In the paper of Hull and Townsend they postulated that the electrically
charged objects with respect to the gauge field $A$ were
charged black holes in $10 - d$ dimensions. This is actually not as strange
as it may seem. Black holes in any dimension
\WaldI\ \WittenI\ with charge $W$
and mass $M$ satisfy the inequality $M \ge \, const \cdot W$. This
looks exactly like our BPS inequality \TypeIIAInequalityI. So, identifying
the electrically charged states  with respect to
$A$ with electrically charged black holes seems rather profitable, and,
as black holes in $10-d$ dimensions also yield black holes in
ten-dimensions,
we can take the electrically charged black holes of $10 - d$ dimensions
to yield electrically charged black holes in ten-dimensions. These
ten-dimensional black holes are the charged objects we have been looking
for. Also, as one may easily see by way of the inequality
$M \ge \, const \cdot W$, the extremal black holes should, in all
likely-hood, be identified with the BPS saturated states.

Now, as we have finally argued for the existence of BPS saturated states
electrically charged with respect to the gauge field $A$, let us see the
promised relation between the central charge $W$ and the electrical
charge with respect to $A$. The derivation is rather simple.
As we previously mentioned, Type IIA supergravity in
ten-dimensions is derivable from eleven-dimensional $N = 1$
supergravity. This eleven-dimensional supergravity theory
possess a supersymmetry algebra with a term of the general form,
\eqn
\ElevenDAntiCommutationRelationI{
	\{ {\cal Q} , {\cal Q} \} \sim {\cal P},
}
\noindent where ${\cal Q}$ is a Majorana supersymmetry charge
in eleven-dimensions and $\cal P$ is an eleven-momentum. As we
noted previously, the ten-dimensional supersymmetry generators
are Majorana-Weyl spinors. So, in dimensionally reducing from
eleven to ten-dimensions the spinor ${\cal Q}$ is split into
two ten-dimensional Majorana-Weyl spinors $Q$ and $Q'$. Also, the
eleven-vector ${\cal P}$ is split into a ten-vector
$P$ and a scalar $P_{11}$. This
process of dimensional reduction leaves us a ten-dimensional
supersymmetry algebra of the general form,
\eqn
\TypeIIAAntiCommutationRelationIV{
	\{ Q , Q  \} \sim P
}
\eqn
\TypeIIAAntiCommutationRelationV{
	\{ Q' , Q' \} \sim P
}
\eqn
\TypeIIAAntiCommutationRelationVI{
	\{ Q_{\alpha} , Q_{\dot \alpha}' \} \sim
	\delta_{\alpha \dot \alpha} P_{11},
}
\noindent exactly the same form as in equations
\TypeIIAAntiCommutationRelationI,
\TypeIIAAntiCommutationRelationII, and
\TypeIIAAntiCommutationRelationIII.
Hence, we see we should, up to a ``constant," make the identification of
$P_{11}$ in \TypeIIAAntiCommutationRelationVI\ with $W$ of
\TypeIIAAntiCommutationRelationIII. As all of this is rather
rough, this identification is up to a ``constant"
which may depend upon the string coupling constant.

As we saw earlier, an electric charge with respect to $A$ corresponds to
a non-zero momentum in the eleventh-dimension,
i.e. $P_{11} \ne 0$. Hence, one may easily
see that the charge with respect to the
gauge field $A$ is indeed the central charge of the ten-dimensional
supersymmetry algebra. Next, let us look at the spectrum of BPS saturated
states which are charged with respect to the gauge field $A$.

As we mentioned earlier, motivated by the U-Duality conjecture we will take
BPS saturated ten-dimensional states charged with respect to $A$
to be charged extremal black holes. Now the
question arises : What is the spectrum of such states? To answer this let us
look at the mass relation for such BPS
saturated states. We have $M = c_{0}W$.
So, for any given charge $W$ there exists a BPS saturated state with mass
given by $M = c_{0}W$. The next question one would ask to find the
spectrum is : What charges $W$ may arise in this ten-dimensional Type IIA
theory? One would assume, in accord with other quantum theories, that
the possible charges are integer multiples of some fundamental unit.
This assumption is also in accord with U-Duality, as U-Duality conjectures a
symmetry which exchanges charges which we know to be quantized
perturbatively with charges with respect to the gauge field $A$. So,
choosing units properly, $W = n_{e}$, where $n_{e}$ is any integer. This seems
the only reasonable point-of-view as the corresponding classical black holes
may take on any charge with respect to the gauge field $A$. Furthermore,
this point-of-view is in accord with our interpretation of
the charge from an eleven-dimensional point-of-view as being the momentum
$P_{11}$ in the compact dimension $S^1$; classically, $P_{11}$ may take on any
value.  So, with this assumption, the spectrum of masses of these BPS
saturated states is given by $M = c_{0} n_{e}$ where $n_{e}$ is an integer.

Next let us consider how BPS saturated states with mass $M = c_{0} n_{e}$
depend upon the string coupling constant which we denote as $\lambda$.
As one may easily find by scaling the metric in equation \TypeIIASupergravityI,
the coupling constant $\lambda$ in Type IIA string theory in ten-dimension
is explicitly given by $\lambda = e^{\phi}$. So, what we want to do is to find
the relation of $\lambda = e^{\phi}$ to the ``constant" $c_{0}$. The easiest
manner in which to do this is to examine the derivation of Type IIA
supergravity from a dimensional reduction of eleven-dimensional supergravity.
As we previously mentioned, dimensional reduction suggests
that we make the identification $c_{0} W \sim P_{11}$. Now, by requiring
the wave function of a particle traveling around the eleventh-dimension
to be single valued we have $P_{11} = n / R$, where $n$ is an integer
and $R$ is the radius of the eleventh-dimension's $S^1$ as measured
by the eleven-dimensional metric $G_{MN} =
e^{A}_{M} e^{B}_{N} \eta_{A B}$. As the dilaton
$\phi$ arises from the eleven-dimensional vielbein, as in equation
\ElevenToTenSupergravityVielbein, one should suspect that the
radius $R$ is related to the value of $\phi$ and this is indeed the
case. The eleven-dimensional metric implies a distance element
$ds^{2} = G^{10}_{mn} dx^{m}dx^{n} + e^{ {4 \over 3} \phi}
( dx^{11} - A_{m} dx^{m})^{2}$,
where $G^{10}_{mn} = e^{a}_{m} e^{b}_{n} \eta_{ab}$ and $A$ and $\phi$
are as they appear in \ElevenToTenSupergravityVielbein.
So, the radius $R$ of the eleventh-dimension's $S^1$  is given by
$e^{ {2 \over 3} \phi}$. Hence, as $W$ is an integer and independent of $\phi$,
one has $c_{0} \sim c e^{ -{2 \over 3} \phi}$,
where $c$ is a true constant. Next we must find the exact relation
between $c_{0}$ and $\phi$ to determine how the BPS saturated states'
spectrum varies with $\phi$ and hence the coupling constant
$\lambda = e^{ \phi }$.

However, this requires a bit of work. Let us start
by explicitly looking at the process of dimensionally reducing
eleven-dimensional
supergravity. Eleven-dimensional supergravity \WittenVI\ has a bosonic
sector which consists of a vielbein $e^{A}_{M}$ and a three-form ${\cal A}_{3}$.
Also, the three-form ${\cal A}_{3}$ gives rise to the field strength
${\cal F}_{4} = d {\cal A}_{3}$. These fields appear in the bosonic
portion of the eleven-dimensional action with the general form \WittenVI,
\eqn
\ElevenDimensionalSupergravityActionI{
	S =
	{ 1 \over 2 } \int \sqrt{ G }
	\left(
		R +
		{ 1 \over 24 } {\cal F}^{2}_{4}
	\right) +
	\int
		{ { \sqrt{2} } \over {3456} }
		\epsilon_{ M_{1} \cdots M_{11}  }
		{\cal F}_{4}^{ M_{1} \cdots M_{4} }
		{\cal F}_{4}^{ M_{5} \cdots M_{8} }
		{\cal A}_{4}^{ M_{9} \cdots M_{11} }.
}
\noindent where $G$ is the determinate of the eleven-dimensional metric,
$R$ is the eleven-dimensional curvature scalar, and
$\epsilon_{M_{1} \cdots M_{11}}$ is the eleven-dimensional totally
anti-symmetric tensor density.

Now, to dimensionally reduce such an action we first must place the
theory on an eleven-dimensional manifold $M \times S^{1}$, with $M$
an arbitrary ten-dimensional manifold. Next we must drop the dependence
of the eleven-dimensional fields on the eleventh-coordinate which
parameterizes $S^{1}$. After this we should decompose the eleven-dimensional
fields into representations of $SO(1,9)$. In particular, this leads to a
decomposition of $e^{A}_{M}$, as in equation
\ElevenToTenSupergravityVielbein,
into a ten-dimensional one-form $A$, a ten-dimensional scalar $\phi$,
and a ten-dimensional vielbein $e^{a}_{m}$.
The three-form ${\cal A}_{3}$ decomposes into a ten-dimensional three-form
$A_{3\, m n p} = {\cal A}_{3 \, m n p}$ and a ten-dimensional two-form
$B_{m n} = {\cal A}_{3 \, m n 11}$. This leads to a ten-dimensional Type IIA
supergravity action of the general form,
\eqn
\TypeIIASupergravityIV{
	S = S_{A} + S_{B}
}
\eqn
\TypeIIASupergravityV{
	S_{A} =
	{1\over 2}
	\int d^{10}x\sqrt{ G^{10} }
	e^{ {2 \over 3} \phi}
	\left(
		R +
		4 ( \nabla \phi )^2 -
		{1\over 4!} F_4'{}^2
	\right)
}
\eqn
\TypeIIASupergravityVI{
	S_{B} =
	-\int d^{10}x \sqrt{ G^{10} }
	\left(
		{ 1 \over 24} e^{- {2 \over 3} \phi} H^{2} +
		{ 1 \over {2 \cdot 2 !} } e^{ 2 \phi} F^{2}
	\right) -
	{1\over 4}\int
	F_4\wedge F_4\wedge B,
}
\noindent where we have employed the notation
of \TypeIIASupergravityI\ for the
field strengths and we have employed the notation $G^{10}_{m n} = e^{a}_{m}
e^{b}_{n} \eta_{a b} $ for the ten-dimensional metric.

Now, one can see, looking at \TypeIIASupergravityI\
and \TypeIIASupergravityIV,
that if we wish to identify the two actions, then we must scale the
ten-dimensional metric $G^{10}_{mn}$ as the powers of
$e^{\phi}$ don't match-up in any easily seen manner. In fact, if we scale as
$G^{10}_{m n} = e^{-{2 \over 3} \phi} g_{mn}$, then we obtain from
\TypeIIASupergravityIV\ an action, in terms of the metric $g_{mn}$, of the
form,
\eqn
\TypeIIASupergravityVII{
	S = S_{NS} + S_{R}
}
\eqn
\TypeIIASupergravityVIII{
	S_{NS} =
	{1\over 2}
	\int d^{10}x\sqrt g e^{-2\phi}\left(R+4(\nabla\phi)^2 -
	{1\over 12}H^2\right)
}
\eqn
\TypeIIASupergravityIX{
	S_{R}=
	-\int d^{10}x \sqrt g\left({1\over 2\cdot 2!}F^2 +
	{1\over 2\cdot 4!}F_4'{}^2\right) -
	{1\over 4}\int F_4\wedge F_4\wedge B.
}
\noindent So, we see that if we identify the metrics of \TypeIIASupergravityI\
and \TypeIIASupergravityVII, then the actions
\TypeIIASupergravityI\ and \TypeIIASupergravityVII\ agree.

Next, let us use this  identification to examine how
the spectrum of BPS saturated states in ten-dimensional Type IIA strings
varies as one varies the coupling constant. As one will remember, the coupling
constant of Type IIA string theory in ten dimensions is $\lambda = e^{\phi}$.
Also, as one will remember, we found that $c_{0}$ of equation
\TypeIIAInequalityI\
is of the form $c_{0} \sim c / R$, where, the radius $R$ was measured in
the eleven-dimensional metric. To obtain a value for $R$ in the new
ten-dimensional metric $g_{mn}$ we must employ the Weyl transformation
$G^{10}_{mn} = e^{-{2 \over 3} \phi} g_{mn}$. Doing so we have $R \rightarrow
R e^{ \phi / 3 }$. Hence, we have $c_{0} = c e^{-\phi / 3} R^{-1}$. Now,
as we previously found $R = e^{ { 2 \over 3 } \phi }$ this implies
$c_{0} = c e^{ - \phi }$. In addition, we have the fact that the Type IIA
string theory
coupling constant in ten-dimensions is $\lambda = e^{\phi}$; so,
$c_{0} = c / \lambda$. This implies that the masses of the BPS saturated states
satisfy $M = c n_{e} /  \lambda$. Hence, we have found how the spectrum of
BPS saturated states depends upon the coupling constant $\lambda$.

Now, as we found in a previous subsection, BPS saturated states of the
above form are neutrally stable against decay; in addition, as the coupling
constant dependence of these states takes such a simple form, they are
neutrally stable for all possible values of $\lambda$. So, in particular
one could consider the step of taking the strong coupling limit
$\lambda \rightarrow \infty$. As the mass spectrum is given by
$M = c n_{e} /  \lambda$, in the limit $\lambda \rightarrow \infty$
one obtains an infinite set of massless states, one for each integer $n_{e}$.
The question is : Can we interpret this $\lambda \rightarrow \infty$
limit as a low-energy field theory of some type?

At low-energies the field theory in question must have an infinite set of
massless states and Type IIA supersymmetry; what might this theory be?
There are only two known consistent theories which
possess Type IIA supersymmetry in ten-dimensions,
Type IIA string theory and Type IIA supergravity.
Neither of these has an infinite set of
massless states at low-energies. So, they are both out. Hence, we still do not
know what low-energy theory might be ``sitting" at the strongly coupled
region of Type IIA string theory. One could postulate that this is some new
ten-dimensional theory, or one could consider that maybe this is the limit of
some previously know or unknown theory in more or less than ten-dimensions.
We will take this latter route as there exists a perfect candidate in
eleven-dimensional supergravity.

As we found earlier, dimensionally reducing eleven-dimensional supergravity
yields a ten-dimensional Type IIA supergravity theory. Also, as we found
earlier, states which are charged with respect to the ten-dimensional
gauge field $A$ correspond to eleven dimensional particles with a
non-zero momentum $P_{11}$. So, it is these states which will appear,
from an eleven dimensional point-of-view, as BPS saturated states. Also,
quite conveniently, there are an infinite number of these states\foot{ As the
momentum $P_{11}$ is in a compact dimension, it is of the form $n / R$,
where $n$ is an integer and $R$ is the $S^{1}$ radius.}. The question
now is : Do all of these states become massless in the correct limit?

To some extent, we have answered this question already. A state with
momentum $P_{11} = n / R$ in the compact dimension $S^{1}$ has mass
$M \sim n / R$. So, in the limit $R \rightarrow \infty$  the masses of
all these momentum states go to zero. Now, does this $R \rightarrow \infty$
limit correspond to the $\lambda \rightarrow \infty$ limit of Type IIA
string theory in ten-dimensions? As we found earlier, as measured in the
ten-dimensional metric $g_{mn}$, the radius $R$ is given by
$R = e^{ \phi } = \lambda$. Hence, the limits correspond
exactly, and we may identify the strong coupling limit of Type IIA
string theory with eleven-dimensional supergravity on a ``large" $S^{1}$.
However, the low-energy limit of M-Theory is eleven-dimensional
supergravity \WittenIII; hence, the low-energy limit of M-Theory is
the strong coupling limit of Type IIA string theory.

%
%

\newsec{ Heterotic $\sim$ M-Theory Equivalence }

In this section we will verify a conjectured relation between M-Theory and
Heterotic string theory. As mentioned in the introduction, the string-string
duality conjecture in six-dimensions conjectures that the strong coupling
limit of a Type II string theory on $K3$ is equivalent to the weak coupling
limit of a Heterotic theory on $T^{4}$. As we showed in the last section,
the strong coupling limit of Type IIA theory in ten-dimensions is the
low-energy limit of M-Theory on a ``large" $S^{1}$. So, along with the
string-string duality conjecture of six dimensions, this implies the conjecture
that the low-energy limit of M-Theory on $K3 \times S^{1}$ with a
``large" $S^{1}$ is equivalent to a weakly coupled Heterotic theory on
$T^{4}$. This conjecture is what we will look at in this section.

%
%

\subsec{ Heterotic String Theory on $T^{4}$ }

In this subsection we will examine the Heterotic string theory on $T^{4}$. But,
let us start the section by looking at the Heterotic theory in its natural
setting, ten-dimensions. As the string-string duality conjecture which we
are examining only relies upon the low-energy limit of the Heterotic theory,
we need only concern ourselves with the low-energy limit of the Heterotic
theory in ten-dimensions. The low-energy limit of the Heterotic theory
in ten dimensions is a $N = 1$ supergravity theory coupled to a $N = 1$ super
Yang-Mills theory \WittenVI. The gauge group $G$ of the
super Yang-Mills theory in ten-dimensions  will only be $G = ( U( 1 ) )^{16}$ as
we will only assume ourselves to be at a ``generic" point in the Heterotic
string theory moduli space. The bosonic field content of the
low-energy effective field theory is as follows :
There is a vielbein $e^{a}_{m}$,
a dilaton $\phi$, a gauge field $A^{I}_{m}$ with
gauge group $G$, and a two-form
$B_{mn}$. The gauge field $A$ and the two-form $B$ give rise to the field
strengths $F = dA$ and $H' = dB$. Also, to write the low-energy
effective field theory we will have need of the three form $H = H' +
\sum_{I} A^{I} \wedge dA^{I}  + {2 \over 3} A^{I}  \wedge A^{I}  \wedge A^{I} $.
With these fields we may write the bosonic portion of the low-energy
effective field theory as follows \WittenVI,
\eqn
\TenDimensionalHeteroticActionI{
	S =
	\int \sqrt{g} e^{ - 2 \phi }
	\left(
		{ 1 \over 2   } R -
		{ 1 \over 4   }  g^{mn} g^{pq} F^{I}_{mp} F^{I}_{nq}  -
		{ 1 \over 4 } ( \partial_{m} \phi )^{2} -
		{ 3 \over 8  }  g^{mn} g^{pq} g^{rs} H_{mpr} H_{nqs}
	\right) ,
}
\noindent where $\partial_{m}$ is the standard ten-dimensional derivative
operator and $g_{mn} = e^{a}_{m} e^{b}_{n} \eta_{ab}$ is the ten-dimensional
metric.

Next our task is to employ the process of compactification to put this theory
on $T^{4}$. This process is similar to that we encountered in dimensionally
reducing eleven-dimensional supergravity to ten-dimensional supergravity.
However, in compactification one keeps the fields' dependence upon the
coordinates parameterizing the compact space. For instance, if we were to
consider compactifying eleven-dimensional supergravity on a $S^{1}$, then
a generic field $K$ would depend on the coordinate $x^{11}$ parameterizing
the $S^{1}$. However, as $S^{1}$ is compact and of radius $R$, we could
employ Fourier analysis to write the field $K$ as follows :
$K = \sum_{ n \in {\bf Z} } K_{n} e^{ i { n \over R} x^{11} }$, where $K_{n}$
is independent of $x^{11}$. Looking at this form of $K$ one can see that the
modes with $n \ne 0$ all have mass $M \sim n / R$. So, one obtains massive
and massless modes upon compactification. This same statement is also true
in the compactification of the Heterotic theory on $T^{4}$. One obtains
massive and massless modes from the same mechanism above. However,
as the string-string duality conjecture which we are attempting to examine
has only need of the low-energy limit of the Heterotic theory, we will
only keep track of the massless modes as the massive ones will be
irrelevant to our inquiry. So, as we are only concerned with the
low-energy limit of the Heterotic theory, our process of compactification
actually reduces to one of dimensional reduction. Hence, in ``compactifying"
we will actually only be dimensionally reducing  the low-energy limit
\TenDimensionalHeteroticActionI\ of the Heterotic string theory in
ten-dimensions.

So, let us start this process of ``compactification." As one will remember,
it consists of three basic steps : Place the theory on the manifold $M \times
T^{4}$, where $M$ is an arbitrary six-manifold. Next, drop the dependence
of the fields upon the coordinates parameterizing $T^{4}$. After this,
decompose the ten-dimensional fields in terms of representations of
$SO( 1, 5 )$. The easiest two portions of this process are putting the theory
on $M \times T^{4}$ and dropping the dependence of the fields upon
the coordinates parameterizing $T^{4}$. These both are simply conceptual
steps and involve no real algebra. Decomposing the fields in terms of
$SO( 1, 5 )$ involves a bit of work.

So, let us consider decomposing the low-energy fields of the Heterotic
theory in ten-dimensions into representations of $SO( 1, 5 )$. Consider
first decomposing the veilbein $e^{a}_{m}$. It isn't that difficult to
figure out the decomposition into representations of $SO( 1, 5)$ upon
looking back at the dimensional reduction of the eleven-dimensional
veilbein \ElevenToTenSupergravityVielbein. One has in this case,
\eqn
\TenToSixSupergravityVielbein{
e^{a}_{m}
\rightarrow \pmatrix{ e^{\alpha}_{\rho} & A^{i}_{\rho} \cr
					          0         & e^{i}_{j}    \cr },
}
\noindent where $e^{\alpha}_{\rho}$  is a six-dimensional veilbein\foot{
The index $\alpha$ is a six-dimensional Lorentz index, the index $\rho$
is a six-dimensional curved index, and the indices $i$ and $j$ go from
$1$ to $4$}, the $A^{i}_{\rho}$ are four six-dimensional one-forms,
$e^{i}_{j}$ are sixteen six-dimensional scalars, and the four-by-six
block of zeros is obtained by making a gauge choice which uses up the
gauge freedom associated with Lorentz transformations between
$M$ and $T^{4}$.

Again looking at \ElevenToTenSupergravityVielbein\ along with
\TenToSixSupergravityVielbein\ one is motivated to interpret the
$A^{i}_{m}$ as a set of four $U( 1 )$ gauge fields. In fact upon
substituting the veilbein \TenToSixSupergravityVielbein\ into
the action \TenDimensionalHeteroticActionI\ one
finds that the one-forms $A^{i}_{m}$ indeed appear in the
six-dimensional action as four $U( 1 )$ gauge fields. Furthermore,
as when we dimensionally reduced eleven-dimensional
supergravity to ten-dimensional supergravity, the gauge transformations of
the $A^{i}_{m}$ correspond to rotations of the appropriate $S^{1}$'s.
The sixteen scalars $e^{i}_{j}$ also have a novel interpretation as
we will see a bit later when we introduce all of the other scalars
present in the ``compactified" theory.

Let us next look at decomposing the one-forms $A^{I}_{m}$. They
give a set of sixteen one-forms $A^{I}_{\rho}$ in six-dimensions as
well as a set of $16 \times 4 =64$ scalars $A^{I}_{i}$.
The set of sixteen one-forms $A^{I}_{\rho}$ are $U( 1 )$ gauge fields
in six-dimensions. The $U( 1 )$ symmetry simply follows from the
sixteen $U( 1 )$'s present in ten-dimensions. The set of sixty-four
scalars $A^{I}_{i}$, as we will see later, have an interesting
interpretation in concert with the other scalars we will reveal in
a moment.

Next let us look at decomposing the two-form $B_{mn}$. It yields
a six-dimensional two-form $B_{\rho \xi}$, four six-dimensional
one-forms $B_{\rho i}$, and six six-dimensional scalars $B_{i j}$.
(Remember $B_{m n}$ is a two-form; thus, anti-symmetric.) One
may explicitly verify that the ten-dimensional action is invariant
with respect to the gauge transformation $\delta B = d \lambda_{1}$,
where $\lambda_{1}$ is a one-form. This gauge transformation also
has a different interpretation in six-dimensions. One has,
\eqn
\BGaugeTransformationI{
	\delta B_{ \rho i } =
	{1 \over 2}
	\left(
		\partial_{\rho} \lambda_{1 \, i} -
		\partial_{i} \lambda_{1 \, \rho}
	\right).
}
\noindent Remembering the fact that one drops the dependence of all
six-dimensional fields upon the coordinates parameterizing $T^{4}$,
this leads to,
\eqn
\BGaugeTransformationII{
	\delta B_{ \rho i } =
	{1 \over 2}
	\left(
		\partial_{\rho} \lambda_{1 \, i}
	\right).
}
\noindent Hence, the four $B_{\rho i}$ may be interpreted as four $U( 1 )$
gauge fields in six-dimensions.

Now let us look at the scalars. All in all we have one $\phi$, six $B_{i j}$,
sixty-four $A^{I}_{i}$, and sixteen $e^{i}_{j}$, a total of eighty-seven scalar
fields. However, these scalar fields admit a six parameter symmetry.
One can consider making a Lorentz transformation on the ``internal"
space $T^{4}$. Such a transformation will only effect the $i$ index of the
field $e^{i}_{j}$. Hence, all of the $e^{i}_{j}$'s are not physical. As the
dimension of this internal $SO( 4 )$ Lorentz group is
${1 \over 2} ( 4 ) ( 4 -1 ) = 6$, only $16 - 6 = 10$ of the $e^{i}_{j}$
scalars are physical degrees of freedom. Hence, instead of eighty-seven
scalars one has only eighty-one. Now there is an interesting manner in which
to write eighty of these eighty-one scalars, the dilaton $\phi$ is the
eighty-first and left out of this construction.

Consider $O(4, 20) / ( O(4) \times O(20) )$. As $O(4,20)$ is of dimension
${1 \over 2} (24) (24 - 1) = 276$ and $O(4)$ and $O(20)$ are of dimensions
${1 \over 2}(4)(4 - 1) = 6$ and ${1 \over 2}(20)(20 - 1) = 190$
respectively, the dimension of $O(4, 20) / ( O(4) \times O(20) )$
is $276 - 6 - 190 = 80$. Equivalent to the number of physical scalars which
we have less the dilaton. Hence, there is the possibility that
one may be able to express
the scalars of the six-dimensional action in terms of an element of
$O(4, 20)$ which is invariant with respect to the action of
$O(4) \times O(20)$. This is indeed the case \SenI. One may put all of
the eighty scalars, excluding the dilaton $\phi$, into a $24 \times 24$
$O( 4, 20)$ matrix $M$ and express the resultant six-dimensional low-energy
field theory in terms of the matrix $M$ in such a way that it is invariant
with respect to the action of $O(4) \times O(20)$.

In fact, let us show that we may actually write the six-dimensional
action in the advertised form. As we found above, the six-dimensional
field theory possess a veilbein $e^{\alpha}_{\rho}$, and hence a
six-dimensional metric $g_{\rho \xi}$, a dilaton $\phi$, a two-form
$B_{\rho \xi}$, and twenty-four $U( 1 )$ gauge fields $A^{I}$, $A^{i}$,
and $B_{\rho i}$. Let us write all of these twenty-four $U(1)$ gauge
fields as $A^{R}_{\rho}$, where $R = 1, \dots, 24$. Also, it possess
eighty scalar fields other than the dilaton. As we mentioned above,
the eighty scalar fields will be packaged in the $24 \times 24$
matrix $M$ which takes values in $O( 4, 20 )$. Upon
``compactification" these fields yield a six-dimensional action
of the general form \SenI,
\eqn
\SixDimensionalHeteroticActionI{
	S = S_{1} + S_{2}
}
\eqn
\SixDimensionalHeteroticActionII{
	S_{1} =
	\int \sqrt{ g}  e^{- \phi}
	\left(
		{1 \over 2} R -
		{1 \over 2} g^{\rho \varrho}
		\partial_{\rho} \phi
		\partial_{\varrho} \phi -
		{1 \over 24} g^{\rho \varrho} g^{\sigma \varsigma} g^{\tau \upsilon}
		H_{\rho\sigma \tau} H_{\varrho \varsigma \upsilon}
	\right)
}
\eqn
\SixDimensionalHeteroticActionIII{
	S_{2} =
		- \int \sqrt{ g}  e^{- \phi}
	\left(
		{1 \over 2} g^{\rho \varrho} g^{\sigma \varsigma}
		F^{R}_{\rho \sigma}
		( L M L )_{R S}
		F^{S}_{\varrho \varsigma} -
		{1 \over 16} g^{\rho \varrho}
		Tr ( \partial_{\rho}M L \partial_{\varrho} M L )
	\right),
}
\noindent where $R$ is the curvature scalar of $g_{\rho \xi}$,
$F^{R} = d A^{R}$, and $L$ is the $24 \times 24$ matrix,
\eqn
\ScalarGroupIdentityMatrix{
	L =
	\pmatrix{ 	-I_{20}	&	0	\cr
			0		&	I_{4}	\cr },
}
\noindent where $I_{n}$ is the $n \times n$ identity matrix. In addition,
the six-dimensional three-form $H_{\rho\sigma \tau}$ is defined by
$H = dB + 2 A^{R} \wedge F^{S} L_{R S}$. So, in deriving
\SixDimensionalHeteroticActionI\ we have accomplished
our goal of finding the low-energy limit of the Heterotic string
theory on $T^{4}$. Now, before we find the low-energy field theory limit of
M-Theory on $K3 \times S^{1}$ let us take a moment to note some properties
of the action \SixDimensionalHeteroticActionI.

Looking at the action \SixDimensionalHeteroticActionI\ one may see
that the vacuum expectation values of the scalar fields $M$ and $\phi$
determine the various coupling constants of the theory. Hence, the various
vacuua are parameterized by the space $O(4, 20) / ( O(4) \times O(20) )
\times { \bf R}$, where a point in the $O(4, 20) / ( O(4) \times O(20) )$
factor is given by the vacuum expectation value of $M$ mod
$O(4) \times O(20)$ and a point in the ${\bf R}$ factor is given by the
vacuum expectation value of $\phi$. Also, note that the gauge
group is given by $( U( 1 ) )^{24}$. Later when we examine M-Theory
on $K3 \times S^{1}$ we will see all of these features reproduced,
evidence that the conjectured relation between Type IIA string theory
and M-Theory is true.

%
%

\subsec{ M-Theory on $K3 \times S^{1}$ }

In this subsection we will examine M-Theory on $K3 \times S^{1}$. But, let us
start by examining M-Theory in its most natural setting eleven-dimensions. The
six dimensional string-string duality conjecture which we are examining only
relies upon the low-energy limit of M-Theory; hence, we need only concern
ourselves with the low-energy limit of M-Theory in eleven-dimensions.
The low-energy limit of
M-Theory in eleven-dimensions is eleven-dimensional supergravity\WittenIII.
Hence, in examining the low-energy limit of M-Theory on $K3 \times S^{1}$
we need only examine eleven-dimensional supergravity
on $K3 \times S^{1}$. So, let us start with an examination of eleven-dimensional
supergravity on $K3 \times S^{1}$.

To start we must compactify eleven-dimensional
supergravity on $K3 \times S^{1}$.
We will do so in two steps. First, we will compactify eleven-dimensional
supergravity on $S^{1}$, then we will compactify the resultant theory
on $K3$ the final result being eleven-dimensional supergravity on
$K3 \times S^{1}$.
So, let us remind ourselves of the eleven-dimensional supergravity action.
The bosonic portion of the action, as we previously presented in equation
\ElevenDimensionalSupergravityActionI, is given by,
\eqn
\ElevenDimensionalSupergravityActionII{
	S =
	{ 1 \over 2 } \int \sqrt{ G }
	\left(
		R +
		{ 1 \over 24 } {\cal F}^{2}_{4}
	\right) +
	\int
		{ { \sqrt{2} } \over {3456} }
		\epsilon_{ M_{1} \cdots M_{11}  }
		{\cal F}_{4}^{ M_{1} \cdots M_{4} }
		{\cal F}_{4}^{ M_{5} \cdots M_{8} }
		{\cal A}_{4}^{ M_{9} \cdots M_{11} },
}
\noindent where the fields are the same as those appearing in
\ElevenDimensionalSupergravityActionI. Next let us compactify
this eleven-dimensional supergravity theory on $S^{1}$. Again, as
we will only be interested in the low-energy limit of the
compactified theory, we will only have need of the compactified
fields which in no way depend upon the coordinate parameterizing
the compact\foot{ As one will remember, such fields have a mass of
order $1 / R$; hence, they are massive and irrelevant to our low-energy
inquiry.} $S^{1}$. As in the case of the Heterotic theory
compactified on $T^{4}$, this means that our compactification
is simply demoted to a dimensional reduction. Thus, we are simply
dimensionally reducing eleven-dimensional supergravity to
ten-dimensions. As one will remember, such a dimensional reduction
leads to Type IIA supergravity in ten-dimensions. Following our
previous results \TypeIIASupergravityIV, one has a ten-dimensional
Type IIA supergravity action of the general form,
\eqn
\TypeIIASupergravityX{
	S = S_{A} + S_{B}
}
\eqn
\TypeIIASupergravityXI{
	S_{A} =
	{1\over 2}
	\int d^{10}x\sqrt{ G^{10} }
	e^{ {2 \over 3} \phi}
	\left(
		R +
		4 ( \nabla \phi )^2 -
		{1\over 4!} F_4'{}^2
	\right)
}
\eqn
\TypeIIASupergravityXII{
	S_{B} =
	-\int d^{10}x \sqrt{ G^{10} }
	\left(
		{ 1 \over 24} e^{- {2 \over 3} \phi} H^{2} +
		{ 1 \over {2 \cdot 2 !} } e^{ 2 \phi} F^{2}
	\right) -
	{1\over 4}\int
	F_4\wedge F_4\wedge B,
}
\noindent where all notation follows that of \TypeIIASupergravityIV.

As we have now compactified M-Theory on $S^{1}$ and taken the low-energy
limit, our next step is to compactify the M-Theory on $S^{1}$ to a M-Theory on
$K3 \times S^{1}$. Or equivalently, as our above compactification is equivalent
to a dimensional reduction, one can consider this process as compactifying
Type IIA supergravity on\foot{ Phrased in this manner one may see that the
relationship of M-Theory to the Heterotic theory in six-dimensions
reduces to the standard string-string duality conjecture \SenI\ in
six-dimensions. So,
at this point we are really finished.} $K3$.

Now, before we put M-Theory on $K3 \times S^{1}$, we must first examine the
process of compactification on more general manifolds than $T^{4}$. In the
case of $T^{4}$  we were rather lucky in that putting the Heterotic theory on
$T^{4}$ the subtitles involved in compactifying a theory were all but invisible.
However, in the case of the manifold $K3$ we must actually work.

Let us start by examining the compactification of a generic $p$-form
field\foot{ Most of this information can be found in standard
references such as Green, Schwarz, and Witten \WittenVI. } $B$.
The standard gauge invariance of a $p$-form field $B$ is given by
the transformation $B \rightarrow B + d \lambda_{ p - 1 }$ where
$\lambda_{ p - 1 }$ is a $( p - 1 )$-form. The gauge invariant field
strength associated with this $p$-form gauge field $B$ is given
by the $( p + 1 )$-form $C = d B$. So, the analog of the Maxwell
action on a manifold $N$ for the $p$-form gauge field is given by,
\eqn
\GenericPFormActionI{
	S( C )  =
	{ { p + 1 } \over { 2 p ! } } \int_{N}
	g^{ a_{1} a_{1}' } \cdots g^{ a_{p} a_{p}' }
	C_{ a_{1} \cdots a_{p} } C_{ a_{1}' \cdots a_{p}' },
}
\noindent where $g_{a b}$ is the metric on the manifold $N$.
One may introduce an inner product on the space of $p$-forms which
is given by,
\eqn
\PFormInnerProductI{
	\langle C , C \rangle =
	{ ( p + 1 ) ! } \int_{N}
	g^{ a_{1} a_{1}' } \cdots g^{ a_{p} a_{p}' }
	C_{ a_{1} \cdots a_{p} } C_{ a_{1}' \cdots a_{p}' }.
}
\noindent So, the inner product is related to the action $S( C )$ as follows :
\eqn
\PFormInnerProductPFormActionRelation{
	S( C ) =
	{ { \langle C , C \rangle } \over { 2 ( p! )^{2} } }.
}
Employing the definition of $C$ allows one to also write the action
$ S ( C ) $ as $S ( C ) = \langle d B , d B  \rangle / ( 2 ( p! )^{2} )$.
From this one may easily see that the equation of motion for the
$p$-form $B$ is given by,
\eqn
\PFormEquationsOfMotionI{
	d^{*} d B = 0,
}
\noindent where $d^{*}$ is the dual of $d$ with respect to the inner product
$\langle \, \cdot \, , \, \cdot \, \rangle$ on $p$-forms.

Consider now if we had some solution $B$ to the $p$-form equations of motion
\PFormEquationsOfMotionI. As the action \GenericPFormActionI\ is gauge
invariant, one may also gauge transform $B$ to $B + d \lambda_{ p - 1 }$ and
obtain a second solution $B + d \lambda_{ p - 1 }$ to the $p$-form equations
of motion. So, to obtain only physical solutions to the $p$-form equations of
motion we need to gauge fix the $p$-form $B$. The $p$-form $B$ has the
gauge freedom $\delta B = d \lambda_{ p - 1 }$; so, we need to impose
enough gauge conditions to fix this symmetry. As $\lambda_{ p - 1 }$ is
arbitrary this corresponds to ${ D \choose {( p - 1 )} }$
gauge conditions if $N$ is
$D$-dimensional. A particularly convenient gauge choice is given by
requiring of $B$,
\eqn
\PFormGaugeCondition{
	d^{*} B = 0.
}
\noindent As $d^{*}$ by definition takes $p$-forms to $( p - 1 )$-forms this
condition represents ${ D \choose {( p - 1 )}}$ constraints, the exact number
we need. We will take this as our gauge condition for $B$.

With \PFormGaugeCondition\ as our gauge condition we may write the
equation of motion for $B$ in a much more conceptually fruitful manner.
The Hodge operator is defined by $\Delta = d d^{*} + d^{*} d$. So, combining
the old equation of motion \PFormEquationsOfMotionI\ with the gauge
condition \PFormGaugeCondition\ we can write the equation of motion
as follows,
\eqn
\PFormEquationsOfMotionII{
	\Delta B = 0.
}
\noindent From a mathematical point-of-view this is very familiar. It is an
equation only satisfied by harmonic $p$-forms. Furthermore, by way of the
Hodge decomposition theorem, $p$-form solutions of
\PFormEquationsOfMotionII\ are in one-to-one correspondence
with the elements of $H^{p} ( N )$, the $p$-th cohomology group
of $N$. Hence, the equations of motion for a $p$-form have gauge
invariant solutions which are given by the harmonic $p$-forms on
the manifold in question and the number of these solutions is the
dimension $b_{p}( N )$ of $H^{p}( N )$ otherwise known as the $p$-th
Betti number.

How does all of this relate to string theory compactifications ? To answer this
question let us consider this formalism on a manifold of the general form
$N = M \times K$, where $K$ is taken to be the ``small" internal manifold and
$M$ is what we would consider as a ``normal" spacetime.  In this case the
Hodge operator splits as follows,
\eqn
\HodeOperatorSplit{
	\Delta = \Delta_{M} + \Delta_{K},
}
\noindent where $\Delta_{M} $ is the Hodge
operator on $M$ and $\Delta_{K}$
is the Hodge operator on $K$. Thus, on the manifold $M \times K$ the
equations of motion for the gauge-fixed $p$-form $B$ take the form,
\eqn
\PFormEquationsOfMotionIII{
	\Delta_{M} B + \Delta_{K} B= 0.
}
\noindent Let us consider what such an equation of motion ``means."

As we have taken $K$ to be the ``small" manifold it has dimensions of order
$R'$ where $R'$ is a length scale much smaller than that we are probing.
So, the eigenvalues of the operator $\Delta_{K}$ are of order $1 / R'{}^{2}$
or a ``large" mass squared.
Looking at \PFormEquationsOfMotionIII\ we see that if $B$ has an
eigenvalue $m^{2} \sim 1 / R'{}^{2}$ with respect to the Hodge operator
$\Delta_{K}$, then on the ``large" manifold $M$ the eigenvalue
$m^{2}$ will be interpreted as giving a mass $m$ to the $p$-form
$B$ on $M$. As $m^{2}$ will be a
``large" mass squared, $p$-forms with non-zero $\Delta_{K}$
eigenvalues will not appear in the low-energy field theory on $M$.
So, when examining the low energy field theory on $M$ resultant
from a manifold of the form $M \times K$, only the zero
eigenvalues of $\Delta_{K}$ will be of any concern.

As we mentioned above in the case of $N$, the $p$-form zero eigenvalues
of the Hodge operator are harmonic forms, and such harmonic forms are
in one-to-one correspondence with the elements of $H^{p}( N )$. This is
also true of the Hodge operator $\Delta_{K}$. The $q$-form zero
eigenvalues of the Hodge operator $\Delta_{K}$ are harmonic forms on
$K$ and in one-to-one correspondence with elements of $H^{q}( K )$.
So, consider a $( p - q )$-form $\alpha$ on $M$ and a $q$-from $\beta$
on $K$. Let us assume that $\beta$ is a harmonic form on $K$ and hence
satisfies $\Delta_{K} \beta = 0$. Together these forms define a $p$-form
$B = \alpha \wedge \beta$ on $M \times K$. If we consider this as a possible
$p$-form gauge field, then the equations of motion
\PFormEquationsOfMotionIII\ for $B$ take the form,
\eqn
\PFormEquationsOfMotionIV{
	\Delta_{N}  \, \, \alpha  = 0.
}
\noindent This is simply the analog of the massless Klein-Gordon equation
in the case of a $( p - q )$-form on $M$. From this construction and the
K\"unneth formula one can see the following general rule : If one has a
$p$-form $B$ before compactification, then it gives rise to $b_{q}( K )$
massless $( p - q )$-form fields upon compactification to $M \times K$,
where $b_{q}( K )$ is the $q$-th Betti number of $K$. We will use this
general rule numerous times as we compactify onto $K3 \times S^{1}$.

Before applying this formalism to the case of M-Theory on
$K3 \times S^{1}$ let us, as a check on the formalism, quickly apply it
to the case of the Heterotic theory on $T^{4}$ to check field content.
By way of the K\"unneth formula and the fact that $H^{0}( S^{1} ) = {\bf Z}$
and $H^{1}( S^{1} ) = {\bf Z}$, one can find the Betti numbers of
the torus $T^{4}$. They are given by $b_{0} = 1$, $b_{1} = 4$,
$b_{2} = 6$, $b_{3} = 4$, and $b_{4} = 1$. The form content of
the Heterotic theory in ten-dimensions is given by $16$ one-forms
$A^{I}$ and a two-form $B$. By way of our ``general rule" the
one-forms $A^{I}$ give rise to $16 \times b_{1}(T^{4})$ zero-forms and
$16 \times b_{0}(T^{4})$ one-forms. Looking at the Betti numbers of $T^{4}$ this
translates to $16 \times 4$ zero forms, which we previously called $A^{I}_{i}$,
and $16 \times 1$ one-forms, which we previously called $A^{I}_{\rho}$.
Similarly, $B$ gives rise to $b_{2}(T^{4})$ zero-forms, $b_{1}(T^{4})$
one-forms,
and $b_{0}(T^{4})$ two-forms. From the Betti numbers of $T^{4}$ this
translates to $6$ zero-forms, which we previously called $B_{i j}$,
$4$ one-forms, which we previously called $B_{\rho i}$, and
$1$ two-form, which we previously called $B_{\rho \xi}$. So,
we can see that this formalism works in the case of the
Heterotic string. Let us consider now applying it to
M-Theory.

In our task of putting M-Theory on $K3 \times S^{1}$, we first put
M-Theory on $S^{1}$ and took the low-energy limit obtaining
Type IIA supergravity. The next step will be to put Type IIA
supergravity on $K3$ and take the low-energy limit. Hence,
to find our final result of M-Theory on $K3 \times S^{1}$ we
need only put Type IIA supergravity on $K3$. Let us now
look at the specifics of this task, much of which is worked
out in \DuffI.

First we will examine the compactification of the forms of
Type IIA supergravity on $K3$. The process of compactifying
the forms of Type IIA supergravity on $K3$ is exactly the general
process we outlined above for a generic $p$-form. So, we will
simply apply this process to each form of Type IIA supergravity.

The Type IIA supergravity theory possess three forms : a one-form
$A$, a two-form $B$, and a three form $A_{3}$. Let us first consider
the one form $A$. The one-form $A$, by way of our general
considerations above, gives $b_{1}(K3)$ zero-forms and $b_{0}(K3)$ one-forms.
The two-form $B$ gives $b_{2}(K3)$ zero-forms, $b_{1}(K3)$ one-forms,
and $b_{0}(K3)$ two-forms. Finally, $A_{3}$ gives $b_{3}(K3)$ zero-forms,
$b_{2}(K3)$ one-forms, $b_{1}(K3)$ two-forms, and $b_{0}(K3)$ three forms.
Now the question is : What are the Betti numbers of the manifold
$K3$? One may find these in the standard reference Griffiths and Harris
\GriffithsI. The Betti numbers of $K3$ are as follows : $b_{0} = 1$,
$b_{1} = 0$, $b_{2} = 22$, $b_{3} = 0$, and $b_{4} = 1$.
So, these imply that the
one-form $A$ gives rise to $1$ one-form. The two-form $B$ gives
rise to $22$ zero forms and $1$ two-form, and the three-form
$A_{3}$ give rise to $22$ one-forms and $1$ three-form. Let us
next consider how to express this field content more mathematically.

Let us work first with the one-form $A$. It gives only one six-dimensional
one-form. Let us denote this one-form as $A'$. As our above
construction of the $p$-form $B$ in terms of the forms $\alpha$
and $\beta$ implies, the one-form $A'$ is related to
the one-form $A$ by,
\eqn
\OneFormRelation{
	A( x^{\rho}, x^{i} ) = A' ( x^{\rho} ),
}
\noindent where we have parameterized $M$ of the splitting
$M \times K3 \times S^{1}$ by $x^{\rho}$ and $K3$ by $x^{i}$.
We may express $B$ and $A_{3}$ in a similar fashion. However,
to do so we must introduce a basis for $H^{2}( K3 )$.

Let us choose an integral basis of the harmonic two-forms on $K3$
and notate this basis as $\omega^{I}$, where, as $b_{2} = 22$, the index
$I$ runs from $1$ to $22$. As $\omega^{I}$
is harmonic, it is closed and co-closed. Hence, ${\hat *} \omega^{I}$, where
${\hat *}$ is the Hodge star operator on $K3$,  is also harmonic\foot{
This follows easily from the fact that up to a sign $d^{*}$ on $K3$ is
$d^{*} = {\hat *} d {\hat *}$ and ${\hat *}^{2} = 1$}. So, as $\omega^{I}$
is a basis of the space of harmonic two-forms on $K3$, there exists a
matrix $H^{I}_{J}$ such that,
\eqn
\HarmoincTwoFormChangeOfBasis{
	{\hat *} \omega^{I} =
	\omega^{J} H^{I}_{J}.
}
\noindent One should note \DuffI\ that the matrix $H^{I}_{J}$ depends
upon the metric on $K3$;  hence, it is a function of the metric's
associated $57$ moduli.

Now, with this basis of harmonic two-forms we may
express the six-dimensional fields which result from the two-form
$B$ and the three-form $A_{3}$. As our above construction of the
$p$-form $B$ in terms of the forms
$\alpha$ and $\beta$ implies, we may employ the basis of harmonic
two-forms $\omega^{I}$ to explicitly write the low-energy fields
resultant from putting M-Theory on $K3 \times S^{1}$. Explicitly,
the two form $B$ gives rise to the two-form $B( x^{\rho} )$ and the
$22$ zero-forms $B_{I} ( x^{\rho} )$ which are related to $B$ as
follows,
\eqn
\TwoFormSpectrum{
	B ( x^{\rho}, x^{i} ) =
	B ( x^{\rho} ) + B_{I} ( x^{\rho} ) \omega^{I} ( x^{i} ).
}
\noindent Similarly, the three-form $A_{3}$ gives rise to the three-form
$A_{3} ( x^{\rho} )$ and the $22$ one-forms $A_{I} ( x^{\rho} )$,
\eqn
\ThreeFormSpectrum{
	A_{3} ( x^{\rho}, x^{i} ) =
	A_{3} ( x^{\rho} ) + A_{I} ( x^{\rho} ) \omega^{I} ( x^{i} ).
}
\noindent Now we are almost ready to write the six-dimensional action
resultant from the low-energy limit of M-Theory on $K3 \times S^{1}$. We
only need to resolve a few more points.

Counting up our one-forms in six-dimensions resulting from this
compactification we have $22$ one-forms $A_{I}$, which come from
the compactification of $A_{3}$, and an additional one-form $A'$,
which comes from the compactification of $A$, for a total of $23$.
However, the Heterotic theory, with which we are trying to match this
theory, has $24$ $U( 1 )$ gauge fields. It looks as if we are one short if we
wish to match these theories up. However, upon a closer look, this is
not the case. $A_{3}$ gives rise to a six-dimensional three-form
$A_{3} ( x^{ \rho } )$. This six-dimensional three-form \DuffI\
gives rise to a six-dimensional field strength
$F'_{4}( x^{ \rho } )$ which is the analog of the ten-dimensional
four-form $F'_{4}$. Upon dualizing this four-form \DuffI\ one obtains
a six-dimensional two-form field strength which yields a six-dimensional
one-form ``vector-potential" which we shall call\foot{ There are several
technical details of this process which we are not covering, but
the interested reader can find them in \DuffI.} $A''$. Hence, in
sum total we have $23 + 1 = 24$ gauge fields each with $U( 1 )$ gauge group,
just the same number of $U( 1 )$ gauge fields which appear in the
Heterotic theory compactified on $T^{4}$.  So, as in the Heterotic theory,
let us place these $24$ one-forms into a convenient package $A^{R}{}$,
where $R$ runs from $1$ to $24$.

Consider next the scalars in the low-energy limit of M-Theory on
$K3 \times S^{1}$. Less the dilaton, we have the $57$ moduli of a $K3$
metric and $22$ scalars $B_{I}$ from the
ten-dimensional two-form $B$. This comes to a total of $79$ one
short of the $80$ present in the Heterotic theory on $T^{4}$.
However, the $57$ moduli present in the metric on $K3$
are not the entire story. There exists \DuffI\  a scalar ``breathing" mode
$\rho$ of $K3$. So, counting this ``breathing" mode, less the dilaton one has
$1 + 57 + 22 = 80$ scalars present in the low-energy limit of M-Theory on
$K3 \times S^{1}$, less the dilaton, exactly the same number present in
the Heterotic theory on $T^{4}$. So, they may indeed be equivalent.

Finally, putting all of this information together, we may write
down a six-dimensional action which results from M-Theory
being compactified on $K3 \times S^{1}$. As in the case of the
Heterotic theory, we may define a $24 \times 24$ dimensional
$O( 4, 20 )$ matrix $M$ which depends upon the $80$ scalars of
the theory mod the dilaton. This matrix $M$ will appear in the
action in such a manner that it is invariant with respect to the
action of $O( 4 ) \times O( 20 )$ so as to yield $80$
physical scalars. Also, in writing the action let us introduce a
prime to all six-dimensional fields to delineate them from the
six-dimensional fields of \SixDimensionalHeteroticActionI. The\break
\eject
\noindent resultant action is \SenI,
\eqn
\SixDimensionalMTheoryActionI{
	S = S_{1} + S_{2} + S_{3}
}
\eqn
\SixDimensionalMTheoryActionII{
	S_{1} =
	\int \sqrt{ g'}  e^{- \phi'}
	\left(
		{1 \over 2} R' -
		{1 \over 2} g^{\rho \varrho}{}'
		\partial_{\rho} \phi'
		\partial_{\varrho} \phi' -
		{1 \over 24} g^{\rho \varrho}{}' g^{\sigma \varsigma}{}'
		g^{\tau \upsilon}{}'
		H_{\rho\sigma \tau}{}' H_{\varrho \varsigma \upsilon}{}'
	\right)
}
\eqn
\SixDimensionalMTheoryActionIII{
	S_{2} =
		- \int \sqrt{ g'}  e^{- \phi'}
	\left(
		{1 \over 2} g^{\rho \varrho}{}' g^{\sigma \varsigma}{}'
		F^{R}_{\rho \sigma}{}'
		( L M' L )_{R S}
		F^{S}_{\varrho \varsigma}{}' -
		{1 \over 16} g^{\rho \varrho}{}'
		Tr ( \partial_{\rho}M' L \partial_{\varrho} M' L )
	\right)
}
\eqn
\SixDimensionalMTheoryActionIV{
	S_{3} =
		- \int
	\left(
		{ 1 \over 8 } \epsilon^{\mu \nu \rho \sigma \tau \epsilon}
		B_{\mu \nu}{}'
		F^{R}_{\rho \sigma}{}'
		L_{R S}
		F^{S}_{\tau \epsilon}{}',
	\right) ,
}
\noindent where $H' = d B'$, $F^{R}{}' = d A^{R}{}'$, the matrix $L$
is the same matrix we encountered in \SixDimensionalHeteroticActionI,
and $\epsilon^{\mu \nu \rho \sigma \tau \epsilon}$ is the totally
anti-symmetric tensor density in six-dimensions.

%
%

\subsec{ Heterotic $\sim$ M-Theory Equivalence }

From the six-dimensional Heterotic action
\SixDimensionalHeteroticActionI\ and the six-dimensional M-Theory action
\SixDimensionalMTheoryActionI\ one may see that the equations
of motion for both theories are equivalent \SenI\  if one makes the
following identifications \SenI,
\eqn
\IdentificationsI{
		\phi' = - \phi
}
\eqn
\IdentificationsII{
		g_{\rho \xi}{}' = e^{- \phi} g_{\rho \xi}
}
\eqn
\IdentificationsIII{
		M' = M
}
\eqn
\IdentificationsIV{
		A^{R}{}' = A^{R}
}
\eqn
\IdentificationsV{
		\sqrt{ g } e^{ \phi } H^{ \mu \nu \rho} =
		{ 1 \over 6 } \epsilon^{\mu \nu \rho \sigma \tau \epsilon}
		H_{\sigma \tau \epsilon}{}'.
}
\noindent The important point to notice about the above identifications
is $\phi' = - \phi$. Upon putting the Heterotic theory on $T^{4}$ the
six-dimensional coupling constant is proportional to $e^{{\phi} \over {2}}$. So,
the weakly coupled region corresponds to $\phi \rightarrow - \infty$.
However, as we found earlier, to identify eleven-dimensional M-Theory
on $S^{1}$ with ten-dimensional Type IIA string theory we must make
the radius of the $S^{1}$ ``large." This, as $R$ is proportional to $e^{\phi'}$,
corresponds to $\phi' \rightarrow \infty$. So, the above identification
\IdentificationsI\ is actually matching the correct limits : The weak
coupling region of the Heterotic theory is matched with the low-energy
limit of M-Theory. Happily this is in accord with our expectations.

%
%

\subsec{ Conclusion }

As it turns out, our derivation is simply a re-statement of the
string-string duality conjecture in a different guise. This could
be seen to be the case when one thought about the fact
that to obtain the low-energy limit of M-Theory on $K3 \times S^{1}$
we simply were placing the Type IIA theory on $K3$. This is the same
exploration which motivated the string-string duality conjecture in
six-dimensions; the fact that the Heterotic theory on $T^{4}$ and
Type IIA theory on $K3$ both have the same moduli space
and have the same equations of motion upon the above
identifications has been known \SenI. So, we have simply found
that the M-Theory $\sim$ Heterotic equivalence conjecture in
six dimensions simply reduces to the string-string duality conjecture
in six dimensions which is ``good" in that it leads to no contradictions,
but is ``bad" in that it leads to no new physics.

A similar argument as we presented in this article also goes through for
the case of the M-Theory $\sim$ Type II equivalence in six-dimensions.
This equivalence places M-Theory on $T^{4} \times ( S^{1} / {\bf Z}_{2} )$
and the Type II theory on $K3$. It equivocates the low-energy limit
of M-Theory on $T^{4} \times ( S^{1} / {\bf Z}_{2} )$ with the weak
coupling limit of the Type II theory on $K3$. This equivalence also
reduces to the standard string-string duality conjecture in
six-dimensions.

One can see this by looking first at M-Theory on $S^{1} / {\bf Z}_{2}$.
The low-energy limit of M-Theory on $S^{1} / {\bf Z}_{2}$ is equivalent
to \WittenIII\  a strongly coupled Heterotic theory in ten-dimensions.
Taking the low-energy limit of M-Theory on
$T^{4} \times ( S^{1} / {\bf Z}_{2} )$ is thus equivalent to taking the
strong coupling limit of the Heterotic theory on $T^{4}$. However,
by way of the standard string-string duality \SenI\ the strong coupling
limit of the Heterotic theory on $T^{4}$ is equivalent to the weak
coupling limit of the Type II theory on $K3$. Hence, the
M-Theory $\sim$ Type II equivalence in six-dimensions
simply reduces to the standard string-string duality conjecture in
six-dimensions. So, again, this is rather ``good" in that it leads to no
contradictions, but is ``bad" in that it leads to no new physics.

\listrefs
\bye